%% file: humscale.tex
\documentclass{acm}

\usepackage{times}  
\usepackage{array, graphics, graphicx, color, xspace, url, multirow, rotating, epsfig, amsmath, subfig}
\usepackage{wrapfig}
\usepackage{tikz}
\usepackage{booktabs}
\usepackage{epstopdf,subfig}
\usepackage{multicol}
\usepackage[format=plain,labelfont=bf,font=small]{caption}
\usepackage{ragged2e,flushend,enumitem,listings}
\usepackage{acronym} 
\usepackage{breakurl}
\usepackage{hyperref, cleveref}
\usepackage{cite}
\usepackage{multirow}
\usepackage[frozencache,cachedir=.]{minted}
\usepackage{appendix}
\usepackage[compact,small]{titlesec}
\usepackage[para]{footmisc}
\usepackage{comment}
\usepackage{paralist}

\usepackage{listings}[mathescape]
\usepackage{color}
\usepackage{outlines}

\hypersetup{pdfstartview=FitH,pdfpagelayout=SinglePage}

\newcommand\paragraphb[1]{\noindent{\bf{#1}}}
\newcommand\paragraphi[1]{\noindent\emph{#1}}
\newcommand\pb[1]{\paragraphb{#1}}
\renewcommand\pi[1]{\paragraphi{#1}}
\newcommand\pghi[1]{\paragraphi{#1}}

\newcommand{\bi}{\begin{itemize}}
\newcommand{\ei}{\end{itemize}}

\newcommand{\ie}{\emph{i.e.,}\xspace}

\newcommand{\eat}[1]{}
\newcommand{\figspace}{\vspace{-10pt}}

\newcommand{\allnotes}[1]{}
\renewcommand{\allnotes}[1]{\textit{#1}} %

\let\svthefootnote\thefootnote
\newcommand\freefootnote[1]{%
  \let\thefootnote\relax%
  \footnotetext{#1}%
  \let\thefootnote\svthefootnote%
}

\graphicspath{{./dia/}}
\begin{document}

\title{Human-Scale Computing: \\ A Case for Progressive Narrow Waist \\ for Internet Applications}

\author{Silvery Fu and Pratyush Das (co-first authors), Sylvia Ratnasamy \\ UC Berkeley}
\maketitle
\begin{abstract}
In the era where personal devices and applications are pervasive, individuals are continuously generating and interacting with a vast amount of data. Despite this, access to and control over such data remains challenging due to its scattering across various app providers and formats. This paper presents Human-Scale Computing, a vision and an approach where every individual has straightforward, unified access to their data across all devices, apps, and services. Key to this solution is the Human Scale Portal, a progressively designed intermediary that integrates different applications and service providers. This design adopts a transitional development and deployment strategy, involving an initial bootstrapping phase to engage application providers, an acceleration phase to enhance the convenience of access, and an eventual solution. We believe that this progressive ``narrow waist'' design can bridge the gap between the current state of data access and our envisioned future of human-scale access.

\end{abstract}

\input{intro}
\input{reality}

\input{overview}
\input{approach}
\input{deploy}

\input{disc}

\onecolumn \begin{multicols}{2}
\bibliographystyle{abbrv} 
\bibliography{humscale}
\end{multicols}

\end{document}

%% file: intro.tex
\section{Introduction}
\label{sec:intro}

With the proliferation of personal devices, web, and mobile applications, individuals generate and interact with diverse forms of data daily. This data, however, is often siloed within the ecosystems created by individual device vendors and app providers, limiting the ability of \emph{users} to access and control their data holistically. The ability to do so is crucial as it enables users and their apps to derive comprehensive insights into their digital lives and actions. Several case studies demonstrate this need, spanning various domains from wearable devices~\cite{fitbit,oura,apple-watch} and Internet of Things (IoT) apps~\cite{smartthings,homeassistant,alexa} to cloud-based services like Dropbox~\cite{dropbox} and Google Drive~\cite{gdrive}. 

This paper envisions Human Scale Computing, a future where every individual has easy and seamless access to all of their devices, apps, and data, all manageable from \emph{trusted human-scale apps} that perform \emph{human-scale (data) access}. For example, imagine a scenario where a user wants to compile a comprehensive health report, drawing data from their fitness tracker, their health app's nutritional information, sleep patterns logged by their smart home system, and their digital medical records. In this scenario, a human-scale app would enable the user to pull all this data together with ease, without having to manually gather and integrate data from each individual source. By doing so, the user could have a holistic view of their health status, potentially leading to more informed health decisions. The human-scale app, in this case, acts as an aggregator, providing human-scale data access that integrates multiple data sources pertaining to and in most cases, owned by the users.

This vision, however, is \emph{not} today's reality. Today, individuals' data is siloed within numerous applications and devices, each with their own proprietary formats and access protocols. Users often struggle with disjointed services~\cite{malki2016data}, inconsistent data formats~\cite{stonebraker2018data}, and a lack of convenient access~\cite{hpi}. To make things worse, many applications currently have minimal incentives to enable user-centric data access and control. These include not just technological challenges, such as API and data heterogeneity, but also commercial considerations. Applications and device vendors often see more value in retaining exclusive control over the data they collect, potentially using it to enhance their services or for targeted advertising.

Is the prospect of Human-Scale Computing overly ambitious? Could this vision be an overreach? One important building block is already in place: regulations like the General Data Protection Regulation (GDPR)~\cite{gdpr} ensure that users have the \emph{right to access} their own data from app providers. The right of access guarantees that personal data must be made available to the user when requested. However, as we will demonstrate through case studies (\S\ref{sec:case}), the right of access alone isn't enough to realize Human-Scale Computing. This is because the right of access doesn't address how \emph{easy} it is to access that data --- not to mention facilitating human-scale access. In other words, the right of access provides the \emph{possibility of access}, but not the \emph{practicality of access} required for our human-scale vision.

How do we transition from the right of access status quo to our end goal of human-scale access? Our key insight, drawing from principles of the Internet, our key insight is the creation of a \emph{narrow waist}, or a standardized common interface. Similar to other interfaces, it needs to be general, user-friendly, and straightforward to maintain. Importantly, considering the reliance of Internet applications, it should be designed for gradual adoption and backward compatibility. This would allow it to integrate smoothly into existing app provider frameworks with minimal disruption, while also accommodating future developments. In essence, we argue for a \emph{progressive} approach to designing this narrow waist.

Specifically, our proposed solution to bridge this divide is the Human Scale Portal (hsPortal), a progressively designed narrow waist that operates in three stages. The first, or the \emph{bootstrapping stage}, is centered around \emph{metadata}. Source app providers, who control the user's data, are requested to share associated metadata - including details on data availability and any required information for data access such as APIs. This information is presented in standardized Data Access Blocks (DABs), which are then cached by the hsPortal and made accessible to human-scale apps (hsApps). Developers can import the DABs from the hsPortal using them in the hsApps. Then, during the \emph{acceleration stage}, we expect source providers to augment the level of detail in the DABs they provide to further simplify data access, possibly in the form of standard query templates for their user's data which hsApp developers can incorporate in the apps. Finally, in the \emph{eventual stage}, we envisage the hsPortal itself evolving to provide data caches and a variety of indexes designed to speed up the query process.

This phased strategy within our progressive design allows for gradual improvements to data access, consistently working towards our end goal of human-scale access. A key aspect of our approach is the \emph{shift in incentives} it can create for applications and device vendors. By adopting this model, vendors can improve their service quality through greater data access, comply with regulations, and attain a competitive advantage. There's also potential for vendors to benefit financially from data sharing, with some of this financial gain possibly passed on to the user. This shift in incentives is intended to promote a more cooperative and beneficial environment that gradually moves towards human-scale access.

In the rest of this paper, we present a reality check on human-scale computing (\S\ref{sec:case}), assumptions of our solution (\S\ref{sec:approach}), including the progressive narrow waist approach (\S\ref{sec:deploy}).

%% file: reality.tex
\vspace{-1em}
\section{A Reality Check for HSC}
\label{sec:case}

\begin{table*}
\centering
\footnotesize
\begin{tabular}{|c|l|c|c|c|c|c|c|c|c|}
\hline
\textbf{\textbf{Domain}} & \multicolumn{1}{c|}{\textbf{\textbf{App}}} & \textbf{Access Method(s)} & \textbf{Export} & \textbf{Search} & \textbf{Analytics} & \textbf{Access} & \textbf{Format} & \textbf{Granularity} & \textbf{Fee} \\ \hline
\multirow{3}{*}{Health} & Fitbit~\cite{fitbitexport} & UI, Programmatic & \checkmark & \checkmark & \checkmark & \checkmark & CSV & Date range & No \\ \cline{2-10} 
 & Oura~\cite{ouraexport} & UI, Programmatic & \checkmark & \checkmark & \checkmark & \checkmark & CSV & Date range & No \\ \cline{2-10} 
 & HealthKit~\cite{healthkitexport} & UI, Programmatic & \checkmark & $\times$ & \checkmark & \checkmark & XML & Date range & No \\ \hline
\multirow{2}{*}{Messages} & Slack~\cite{slackapi, slackexport} & Programmatic & \checkmark & \checkmark & $\times$ & \checkmark & JSON & Date range & Yes \\ \cline{2-10} 
 & Whatsapp~\cite{whatsappexport} & User & $\times$ & $\times$ & $\times$ & \checkmark & TXT & Full history & No \\ \hline
\multirow{2}{*}{Social} & Twitter~\cite{twitterexport} & UI, Programmatic & \checkmark & \checkmark & \checkmark & \checkmark & JSON, HTML & Date range & No \\ \cline{2-10} 
 & Facebook~\cite{facebookexport} & UI & \checkmark & \checkmark & \checkmark & \checkmark & JSON, HTML & Full history & No \\ \hline
\multirow{3}{*}{\begin{tabular}[c]{@{}c@{}}Personal \\ Finance\end{tabular}} & Paypal~\cite{paypalAPI, paypaldataaccess} & User, Programmatic & \checkmark & \checkmark & $\times$ & $\times$ & CSV & Date range & No \\ \cline{2-10} 
 & Cashapp~\cite{cashappexport} & UI & \checkmark & \checkmark & $\times$ & $\times$ & CSV & Date range & No \\ \cline{2-10} 
 & Amex~\cite{amexhistory} & UI & \checkmark & \checkmark & $\times$ & $\times$ & CSV, XLS & Date range & No \\ \hline
\multirow{2}{*}{IoT} & Alexa~\cite{amexhistory} & UI & $\times$ & $\times$ & $\times$ & $\times$ & Interface & Date range & No \\ \cline{2-10} 
 & Samsung~\cite{smartthingsCLI, smartthingsGDPR} & UI, Programmatic & \checkmark & $\times$ & $\times$ & $\times$ & CSV & Full history & No \\ \hline
\end{tabular}
\caption{{\bf Comparison of personal data access methods across applications and application domains.}}
\label{tab:comparesources}
\figspace
\end{table*}

\subsection{Regulation}
\label{subsec:regulation}
The EU’s introduction of GDPR in 2016 established legal definitions of data privacy for citizens of the European Union and guidelines that data controllers had to obey when storing and operating on customer data. The GDPR tenet of “right of access” to personal data would allow for users to have the legal right to access their personal data stored in the data silos of consumer data applications and to request for this data for their own use and to understand how the application is using their data. In this section, we refer to the pertinent GDPR articles to understand the current state of compliance with right of access for GDPR and introduce current solutions to establish a GDPR-compliant infrastructure and their potential shortcomings.

According to GDPR Article 15~\cite{gdpr-art15}, right of access is the ability for a data subject to make a request to a data controller to check if their personal data is present in the controller’s data silo, and, if it is present, should be provided with any personal data which has been connected concerning them and information about processing purposes, categories of personal data processed, and planned duration of storage. The right of access must be easy and free to exercise and data must be provided within one month of the request. This regulation presents the opportunity for users to gain an understanding of their data and develop personal applications on their data, rather than having their data sit within silos as training data for corporations. Despite this newfound access for users, however, the processes for accessing personal data remain complex, hindering users from exercising their right of access as we'll discuss next.

\subsection{Status Quo}
\label{subsec:quo}

Finding a solution for the restricted ability for users to access their data requires an understanding of the current workflows of personal data access. We studied the data access workflows of popular consumer data applications to discover the variations in data access methods between industries and applications. To properly  identify these variations, we formed the following classifications for the workflows:
\begin{inparadesc}
    \item[user interface access] is a portal through which a user can directly request for personal data,
    \item[user access] is when a user has to email or call the company for access to data, and
    \item[programmatic access] is the ability for users to access their data through API call.
\end{inparadesc}
We then went through the process of trying to access the data through the user portal and developer environment. The study results are compiled in Table ~\ref{tab:comparesources}.

Each of these applications has a public statement of GDPR compliance and meets the minimum threshold of having some workflow in which a user can make a request for their own personal data and receive their data that the controller is holding. However, some apps charge fees for data access or limit the exporting of data to only between their own apps, such as with Apple limiting access to health data it collects from users. Messaging and social network apps, such as Slack, charge a fee for programmatic access to data, while some apps, such as Whatsapp, lack a programmatic interface altogether. Other applications, such as banking apps, do not provide programmatic access to data due to data security and access control fears. 

Meanwhile, heterogeneity of data formats of personal data export and the subsequent data engineering disincentives development of apps that pull user data from Internet apps. The permutations of data integrations that arise when developing a personal data app (e.g. mapping resting heart rate measured by a wearable to caloric intake measured by a food logging app) increases the cost, complexity, and maintenance responsibility for developers. Through this heterogeneity, there is immense demand by developers for integrator-aggregator third-party APIs, such as Plaid or Mint, which develop and maintain pipelines from the silos of the source applications and pull values from them into a unified data structure that would be easier for developers to build applications with. However, because integrator-aggregator systems operate as businesses, developers take on immense expenses to build applications. The restriction to development through API usage fees creates greater friction in the data access process, as users cannot programmatically operate on their own data for free. While the above deficiencies were domain and application specific, the reviewed applications collectively failed to meet standards of data access methods and capabilities required to enable users to take advantage of right of access for human-scale access.

\subsection{Existing Approaches}
\label{subsec:prior}

The study allowed us to identify and develop the requirements for a solution to the problem presented by the lack of an easy-to-implement user and programmatic interface for the right of access. We have determined any potential system that enables this interface must contain the following traits:

\pb{Ease-of-Use}: Every source application must come with an intuitive user interface to allow users to easily request access for their data, allowing those without programming experience to perform basic queries on their personal data.

\pb{Ease-of-Development}: Every source application must provide programmatic access through API or other methods for developers to easily programmatically query upon their data stored within the source application’s data silos and operate upon their personal data as they please.

\pb{Generality}: Responses to data requests from a category of data silos must be uniform for all source applications within that category. For example, all wearable computing source applications would follow the same data schema for data requests. This allows for human-scale application developers to create applications that can scale without extensive integration development.

\pb{Deployability}: We define deployability as the speed and predictability with which a solution can be deployed into existing data workflows for source applications and hsApp developers. Any solution that enables human-scale computing should be mostly compatible with the existing data infrastructure of source applications and allow for an easy onboarding experience for all stakeholders (source applications, hsApp developers, hsApp users).

\pghi{(1) Human Programming Interface} project \cite{hpi} is a set of open source modules and libraries that utilizes a combination of data mirroring (i.e. storing a record of online data on a local file) and local data integrations (i.e. writing a script to parse values out of data returned from API calls) for easier search and analytics on personal data. While data mirroring allows for faster discovery and search on personal data and is deployable as a user facing API, it takes up storage from the user and requires consistent merging of data to maintain a single source of truth. Developing a class of functions to access data from every single source app would only reduce the integration time per source application, but does not reduce the complexity of querying and consolidating data from a multitude of sources in a development environment.

\pghi{(2) GDPR-compliance-by-construction} is a conceptual backend that “allows users to seamlessly introduce, retrieve, and remove their personal data without manual labor on the developer’s part” \cite{schwarzkopf2019position}. In the backend of each source application, there are shards for each user, where all data associated with the user is stored. Outside systems will then query from the shards for processing, with the user maintaining control over the use of their shard. Though effective in allowing the user to maintain control of their data, compliance-by-construction requires a complete overhaul of existing database systems and enables developers to easily interface with multiple source applications. 

\pghi{(3) Local-first software} is a set of principles for software that involves storing all changes to data locally and keeping a cluster of devices that can see and accept those changes to data, as a local version of a cloud storage drive, with the changes syncing once the devices are on the same network \cite{kleppmann2019local}. While local-first would be effective for content development applications and provides a multi-user interface while keeping data changes local and private to a user, developers would have to overhaul their data infrastructure to support a local drive for each user, rather than pull from a centralized database.

\begin{table}
\centering
\footnotesize
\begin{tabular}{|c|c|c|c|c|}
\hline
\textbf{System} & \textbf{DMR} & \textbf{GCBC} & \textbf{LFS} & \textbf{HSC} \\ \hline
\textbf{Ease-of-use} & \checkmark & \checkmark & $\times$ & \checkmark \\ \hline
\textbf{Ease-of-dev.} & \checkmark & $\times$ & - & \checkmark \\ \hline
\textbf{General} & $\times$ & $\times$ & $\times$ & \checkmark \\ \hline
\textbf{Deployability} & $\times$ & $\times$ & $\times$ & \checkmark* \\ \hline
\end{tabular}
\caption{\textbf{Comparison of human-scale access with existing solutions.} DMR: Data mirroring. GCPC: GDPR-compliance-by-construction. LFS: Local-first software. HSC: Human-scale computing.}
\label{tab:comparesolutions}
\figspace
\end{table}

\subsection{Takeaways}
\label{subsec:takeaway}
The minimum bar for right-of-access defined in the GDPR legislation does not address the difficulty for users to access a specific piece of data through an easy workflow that enables the right-of-access to be exercised. We believe that a lack of such an interface has led to a stagnancy of valuable data, a reduced ability for data subject’s to perform analysis on their own data, and for meaningful consumer data applications to be developed, representing an overall loss of value of data to those who produce it. We call this desired ability for users to easily interface with their personal data stored by Internet applications as \textbf{Human Scale Computing} (HSC). The goal of HSC is to address the problems created for users by a lack of an intuitive data interface and enables a new paradigm of development in which developers can build applications that scale user access to personal data to all data subjects. We describe HSC in the following section.

%% file: overview.tex
\section{Overview of HSC}
\label{sec:overview}

Having established the problem and describing the limitations of current solutions, we will describe the assumptions, use cases, and goals we have in mind while developing HSC. We establish the following terminology to describe the various part of our infrastructure: 
\begin{inparadesc}
\item[user] is a data subject that wants to gain access to personal data through HSC, 
\item[source app] is an Internet application which a data subject provides data to/a data controller collects data from, \item[human-scale application (hsApps)] is an Internet application developed for HSC that enables human-scale access to data, \item[data access block (DAB)] is a programmatic set of instructions on how a users can access pertinent data, and \item[hsPortal] is an interface that allows users to grant authorization to hsApp developers to query on their data and employs a narrow waist to provide general DABs for each source application domain.
\end{inparadesc}

\pb{Assumptions:}
Implementing hsPortal assumes participation by data subjects, data controllers, and developers through provision of credentials, programmatically defined data access methodology, and refactoring of development to integrate within the hsPortal data architecture. We assume that data controllers are willing to provide metadata, specifically a programmatic method of data access, the fields of data pertinent to the request, and the structure of the data within a query response. We assume that the narrow waist is provided the instructions required to restructure the provisioned data access method into a general schema for developers. We assume that developers will integrate the programmatic DAB within their data retrieval workflows that allows for efficient data retrieval across data sources (for example, a developer makes a single data request to the hsPortal for wearable computing data, rather than to multiple individual wearable computing silos). Finally, we assume that there will be no procedures in hsPortal which reduce the integrity of a subject’s data or allow a developer to view the exact subject from which the data came from. The data will be assumed to be pseudonymized and hsApp backends will not require inspection of the data subject’s identity. 

\pb{Use Cases:}
The primary use cases of hsPortal are as a developer interface to enable developers to build human-scale applications and as a user interface allowing data subjects to maintain control of and access to their data. The benefits for developers building on hsPortal is the ability to write software logic once and build using only general integration with the hsPortal, while maintaining accuracy and functionality of data requested directly from a source application silo. This greatly reduces the number of data integrations a developer has to account for when building consumer applications that integrate data from multiple source applications. We envision examples of human-scale applications to include search engines for personal data, health analytics dashboards visualizing data from a multitude of health tracking devices and applications, and messaging applications that can provide context on conversations from a multitude of previous conversations in separate channels. To use hsApps, users only need to specify their source app credentials in a user-facing portal, and to develop hsApps, developers only need to request data from hsPortal. This simplicity incentivizes participation by both developers and users, which in turn will incentivize cooperation by data controllers to help scale usage and functionality of hsPortal for all stakeholders. 

\pb{Goals:}
We establish the following goals for the system:

\noindent
Goal \#1: The hsPortal should assist developers in sourcing data for hsApps by providing the appropriate data controller DAB when the developer makes a query for user personal data. The data sourcing functionality should increase the productivity of developers in discovering which data controlling applications the user utilizes for a specific functionality and then return the appropriate user data for accessing data within that application.

\noindent
Goal \#2: A DAB returned by the hsPortal should contain information on the existence of the user’s data within the data controller’s data silo and instructions for accessing this data. 

\noindent
Goal \#3: The hsPortal should return a DAB that enables developers to query on data within a source app's silo. In eventual deployments, the hsPortal should have capability to perform federated search on multiple data silos and should return user data values in a uniform data schema, rather than just metadata. 

\noindent
Goal \#4: The hsPortal should be interfaced by developers in an intuitive programmatic fashion such that the output of a query to the hsPortal can be easily integrated within a human-scale application. The hsPortal should also be easily accessible for data controllers to provide the required metadata and/or data for DAB population and provision. 

%% file: approach.tex
\section{Designing HSC}
\label{sec:approach}

HSC seeks to accomplish Goal \#1-\#4 which requires a concrete set of operations to guide development. We outline the three basic operations supported by hsPortal to enable HSC and then briefly sketch the eventual solution.

\subsection{Operations of the hsPortal Narrow Waist}
\label{subsec:operations}
The hsPortal employs a progressive narrow waist to assist developer workflows. The function of the narrow waist is to take in a query from a user and return a programmatic DAB for the user to actually perform the query with. The DAB must be general and uniform across data sources and must enable developers to query and ingest data without designing multiple integrations for a single data type. The narrow waist is able to generate DABs through the following operations:

\pb{Separation of data and metadata}: To store and process all user data would require computational and storage capabilities that reduce the advantages of using a narrow waist for both data controllers and users. Therefore, to achieve efficient information retrieval, the narrow waist separates the view of metadata (information about presence, structure, format of data) and data (values stored within the source application silos). This separation is done by the data controlling entities, who determine the required metadata for data access and develop a set of templates for metadata requests that they can fill with user credentials and then send back to the developer as a DAB through hsPortal.

\pb {Exposing the data access method}: The metadata DAB will contain a programmatic method on how the human-scale application can access the requested user data. While the DAB contains metadata, this method should be returned as an executable script that the application can process in order to query on user data.

\pb {Simplifying the data access process incrementally}: Integration of the metadata DAB within the user data workflow should be intuitive for developers. Initially, the script should execute the data request from the data controller silo and return the data in its raw format, while including programmatic instructions on the specific fields in which the requested data lies, so that parsing can be automated by the human-scale application and placed in a general format on the application side. An eventual solution will perform the operations of the script and perform the parsing required within the HSC infrastructure to get the data into a general format that the developer can use immediately.

\subsection{An Eventual Solution}
\label{subsec:ideal}
An eventual narrow waist would change the way developers structure their data flows in human-scale applications, with the hsPortal acting as a single source of truth for developers to query from without having to account for changing data structures, versioning, and dependencies of integrations, and focus on developing interfaces with the acquired general data format. The ultimate goal for hsPortal would be to act as the backend for all data applications that fall within our definition of a human-scale application. In this subsection, we will describe the eventual flow for users and developers to use the hsPortal for human-scale access to data.

\pb{User Workflow}: To enable the flow of data between source application and hsApps, users must have a way to authorize hsApps to query on their data stored in source applications. We envision that a single user would have designated applications for each aspect of their life. For example, a user might designate Fitbit and Oura as their wearable devices, MyFitnessPal for diet-tracking, and PayPal for online banking. We first propose the design of the hsPortal to implement a single-sign on (SSO) interface. This would allow users to sign into their various applications, specify which applications they use for various functions of their lives, and provide authorization to the hsPortal to facilitate the flow of data between these source applications and third-party developers. Users would only have to perform this sign-on once per application. Authorization would mean the transferring of the credentials required to make a request for data,which is enabled in most standard SSO platforms. A user would then be able to authorize hsApps to utilize their personal data upon download of these applications. Within the hsPortal, users would have the ability to control data access of hsApps and change source applications for their functions.

\begin{figure}
    \centering
    \includegraphics[scale=0.6]{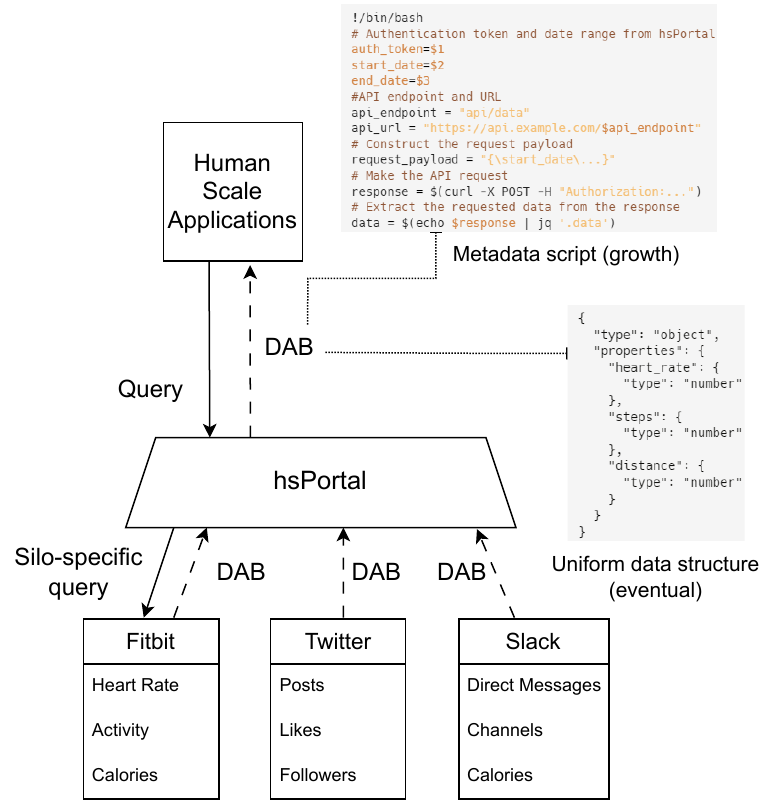}
    \caption{HSC architecture with data flow between source and hsApps.}
    \label{fig:humscalearch}
    \figspace
\end{figure}

\pb{Developer Workflow}: A developer wanting to develop a human-scale application would only have to interact with the hsPortal to create their user data views within the hsApp. For each hsApp, a developer would have to define the specific data functions they would need data from and conditions for the data they would want to work with. These set of conditions will define the fields of user data that the hsApp will query upon. Similar to the module proposition in the Human Programming Interface \cite{hpi}, a developer would be able to create a user object and be able to query general human function data from that user object using a predefined set of functions to access that data within the user object. This data will be general to all data sources related to that function.  In an eventual solution, the result of this query will be a single block of data that the developer can operate upon. This block of data will be structured by a set of schema that best captures the fields of as many source applications as possible, while maintaining a high degree of granularity per source application. In general, these hsApps will not be designed to favor the features of one source application over another, so there should be no disruption to data operations by using a predefined schema, similar to that in Figure~\ref{fig:humscalearch}.

\pb{Data controller workflow}: As a data controller wanting to participate in human-scale computing, the main responsibilities are to allow hsApps to make free data queries on the your data silos where the user data is stored, to provide query templates for the various queries that a hsApp could make, and to provide access to a general mapping of where the schema fields are present within the raw returned data structure, most of which is available through a metadata view of the data controller’s silos. These query templates are generally available in most developer documentation, so provision of a query template should be a low hanging fruit, and developers should only have to provide the parameters to make a request for data their application requires (not including any user specific parameters, which the hsPortal shall populate for each specific hsApp user). In the eventual solution, the narrow waist will populate the query template with the required parameters, make the request to the source application silo, collect the returned data, populate the general data block using the mapping provided by the data controller, and return the DAB to the hsApp.

%% file: deploy.tex
\subsection{A Progressive Deployment for HSC}
\label{sec:deploy}

We propose a methodology termed Progressive Narrow Waist (PNW) for the hsPortal, inspired by recent work on deploying Internet scale services for data revocation~\cite{tet}. PNW begins with a transitional, easy-to-deploy design that doesn't require immediate buy-in from dominant industry entities. While the initial design isn't fully scalable, it aims to shift user and societal expectations, ultimately motivating these industry incumbents to see the value and adopt the approach we advocate. The intellectual challenge in implementing PNW lies in identifying transitional designs that, firstly, engage willing stakeholders to initiate deployment, and secondly, can effectively shift incentives for larger industry entities, motivating them to embrace the system proposed.

\pb{1. Bootstrapping phase: engaging source apps.} The progressive design of our narrow waist starts with this stage. The bootstrapping phase serves as the transitional design, presenting an initial version that stakeholders can readily deploy. At this juncture, it's expected that data controllers may have little visibility or motivation to implement the hsPortal within their systems. Users would have an interface to securely authorize hsApp developers to execute queries on their data silos, with the data pseudonymized to protect user identities. The developers can query specified data silos through the hsPortal, examine the metadata DABs of the response, and design general schema capturing the commonly requested user data fields. While this phase involves data engineering, it allows developers to conduct efficient queries and introduces metadata DABs into the development process.

\pb{2. Growth phase: expanding capabilities.} Following the bootstrapping phase, the growth stage expands the system capabilities, potentially shifting industry incentives and encouraging adoption by larger entities. The hsPortal evolves to allow users to specify the source applications they use for various functions, facilitating the interface to associate these applications with the data fields requested by an hsApp. When a request is made, the hsPortal performs a federated query over the associated data silos and returns executable metadata DABs for each of them. These DABs, when executed, would perform queries on silos for the requested user’s data. The developers can then integrate these executable DABs into their applications, populating the general schemas of data with data from the responses.

\pb{3. Eventual solution: realizing human-scale access.} The final stage of our design, \ie the eventual solution discussed in \S\ref{subsec:ideal}, brings the complete vision of human-scale data access to fruition. Specifically, when the hsApp executes a federated query over data silos, the hsPortal populates a query template for each data silo and performs the query for user data directly. It then automatically populates the data schemas associated with the queries with the values received from the federated queries. At this point, hsApp developers only need to interact with the general data structures containing user data, rather than having to execute a script or perform data engineering to achieve generality.

%% file: disc.tex
\section{Future Research Directions}
\label{sec:disc}

\pb{Improving metadata standards.} HSC relies on metadata from source applications to form DABs, which serve as a bridge between data silos and human-scale apps. This prompts further research into how we might improve metadata standards. Effective metadata standards need to be expressive enough to represent the necessary information about data from various sources~\cite{balaji2016brick}. Besides making these standards more flexible and comprehensive, an interesting direction is to explore how to how to automate the creation and updating of metadata, thereby reducing the burden on source apps and ensuring DABs remain accurate and up-to-date.

\pb{Incentivizing data sharing.} While HSC indicates a positive step toward more cooperative data sharing practices, we envision future research could focus on exploring additional mechanisms to incentivize data sharing. This might involve developing new business models, exploring potential policy and regulatory interventions~\cite{oecd}, and understanding how different incentives might influence application providers' behavior. Ultimately, the goal is to create a data sharing ecosystem where providers see the benefits of participating, thereby contributing to the realization of human-scale access.

\pb{Enhancing user-centric data control.} Central to our proposal is the idea of enabling users to exercise greater control over their data. By presenting users with a unified interface to manage and access their data across multiple sources, the HSC seeks to enhance user-centric data control. However, realizing this vision also raises several research questions. For example, how can we ensure that users truly understand the implications of their data control choices? How can we design the human-scale apps interface to be user-friendly, intuitive, and transparent, so users can make informed decisions about their data? Answering these questions might involve user experience studies, designing user-friendly data visualization tools for data from different app domains, and applying privacy-preserving techniques that can balance the additional data access capability with data protection.